# The changing role of cited papers over time: An analysis of highly cited papers based on a large full-text dataset

Gege Lin

College of Economics and Management, Beijing University of Technology, China

lingege@bjut.edu.cn

https://orcid.org/0000-0003-0137-1471

Nees Jan van Eck

Centre for Science and Technology Studies (CWTS), Leiden University, The Netherlands

ecknjpvan@cwts.leidenuniv.nl

https://orcid.org/0000-0001-8448-4521

Haiyan Hou

School of Public Administration and Policy, Dalian University of Technology, China

houhaiyan@dlut.edu.cn

https://orcid.org/0000-0002-2142-4457

Zhigang Hu

Institute for Science, Technology and Society, South China Normal University, China

huzhigang@scnu.edu.cn

https://orcid.org/0000-0003-1835-4264

# Abstract

This paper examines how the role of cited papers evolves over time by analyzing nearly 900 highly cited papers (HCPs) published between 2000 and 2016 and the full text of over 220,000 papers citing them. We investigate multiple citation characteristics, including citation location within the full text, reference and in-text citation types, citation sentiment, and textual and bibliographic relatedness between citing and cited papers. Our findings reveal that as HCPs age, they tend to be cited earlier in papers citing them, mentioned fewer times in the full text, and more often cited alongside other references. Citation sentiment remains predominantly neutral, while both textual and bibliographic similarity between HCPs and their citing papers decline over time. These patterns indicate a shift from direct topical and methodological engagement toward more general, background, and symbolic referencing. The findings highlight the importance to consider citation context rather than relying solely on simple citation counts. Large-scale full-text analyses such as ours can help refine measures of scientific impact and advance scholarly search and science mapping by uncovering more nuanced connections between papers.

# 1. Introduction

Citation analysis is widely employed in research evaluations around the globe, influencing the allocation of billions of dollars in funding and impacting the work and career of millions of researchers. Addressing any associated issues or limitations is therefore essential (e.g., Zhao & Strotmann, 2020). A longstanding criticism of citation analysis practices is that citations to cited works are all treated equally, regardless of their significance to the citing paper (e.g., Aljuaid et al., 2021; Zhu et al., 2015). Papers are cited for diverse reasons, and often only part of the cited references are essential to the citing paper, while others are more perfunctory. Consequently, simple citation counts fail to capture the underlying intentions and motivations behind citations (e.g., Lyu et al., 2021; Vyas et al., 2020).

By leveraging the full text of citing papers when performing citation analysis, the contribution of the cited paper to the citing paper could be evaluated in a more effective and reasonable way. Previous research has for instance already focused on the fact that citations can appear in different sections in the full text of a paper (e.g., Thelwall, 2019a), be mentioned once or multiple times (e.g., Ding et al., 2013; Hu et al., 2017), include one or multiple references (e.g., Lin et al., 2019; Lyu et al., 2021), and be given with different sentiments (e.g., Yousif et al. 2019).

While previous research has examined citation location, citation type, and citation sentiment, these studies typically rely on relatively small full-text datasets covering short time spans. Moreover, as papers age, their role and the reason for citing them may change. Despite this, the impact of these changes on citation patterns within the full text of papers has rarely been investigated.

To address these gaps, this study uses a large, disciplinary broad full-text collection to analyze how papers are cited in the full text of their citing papers and how these citation patterns change over time. We focus on nearly 900 highly cited papers (HCPs) published between 2000 and 2016 and examine the full text of more than 220,000 papers citing them. Our analysis is guided by the following research questions (RQs):

- RQ1: When a paper is cited, how does the citation location change over time? Will the paper be cited more often in the beginning, middle, or end of the full text of the citing paper?

- RQ2: When a paper is cited, how does the cited reference type change over time? Will the cited reference of the paper increasingly be mentioned once or multiple times in the full text of the citing paper?

- RQ3: When a paper is cited, how does the in-text citation type change over time? Will the paper increasingly be cited alone or together with other references in the full text of the citing paper?

- RQ4: When a paper is cited, how does the citation sentiment change over time? Will the surrounding text become more positive, negative, or neutral?

- RQ5: When a paper is cited, how does the degree of relatedness between the citing paper and the cited paper change over time? We consider two types of relatedness, namely textual relatedness based on the similarity of titles and abstracts and bibliographic relatedness based on the number of shared cited references.

By addressing these questions through large-scale full-text analysis and examining multiple aspects, we characterize the evolving role of cited works over time.

The remainder of this paper is structured as follows. The literature review section revisits related studies in citation analysis. The data and methods section describes the performed data acquisition, data processing, and data analysis. The key findings are reported in the results and discussion section. The final section contains conclusions, limitations and suggestions for future work.

# 2. Background and related work

This section explores the evolution of citation analysis, transitioning from simple citation counting to more nuanced approaches that analyze the functions and contexts of citations. We begin by tracing the shift from static citation metrics to dynamic, context-sensitive citation analyses, highlighting the limitations of relying solely on citation counts and discussing the need for more detailed in-text citation analysis. Next, we examine the growing understanding of citing behavior, emphasizing the importance of authorial intention in interpreting citations and recognizing the diverse motivations behind them as crucial for accurately assessing the impact and influence of cited works. Finally, we synthesize scholarly perspectives on HCPs, discussing how their functions and the motivations for citing them can shift over time.

## 2.1 From counting citations based on reference lists to in-text citation analysis

Gross and Gross (1927) is one of the foundational papers in the field of citation analysis. It is one of the first studies that demonstrated the use of citation counts to determine the importance of scholarly works. This approach laid the groundwork for later developments in citation analysis and the construction of bibliographic databases such as Web of Science and Scopus that index citations. Although bibliometric indicators used in citation analysis have become more advanced over time, the basic principle of measuring the impact of a scholarly paper has not changed much. It typically involves counting the number of times the paper is referenced in the bibliographies or reference lists of other papers (Garfield, 1955). A higher citation count is typically an indication of greater influence and visibility (e.g., Aksnes, 2003; Bornmann, 2014; Tahamtan & Bornmann, 2019).

Counting citations from reference lists is a very rudimentary way to get an impression of the use and importance of a scholarly paper. It overlooks important factors, such as the significance of the paper to the works citing it and the context in which it is referenced. Recognizing that not every citation has the same meaning and value, researchers started to discuss and study the use of in-text citations (e.g., Ding et. al, 2014). In-text citations are the references in the body or full text of a paper. Some early studies in this direction are, for instance, Cano (1989), Maričić et al. (1998), McCain & Turner (1989), and Voos & Dagaev (1976). Access to the full text of scholarly papers in machine-readable formats has made it possible to study in-text citations in greater detail. This has enabled studies focusing on how citations are distributed throughout the full text (e.g., Boyack et al., 2018; Thelwall, 2019a; Zhao & Strotmann, 2020), the number of times a particular work is cited within a single paper (e.g., Ding et al., 2013; Hu et al., 2017; Pak, Wang, & Yu, 2020; Zhao et al., 2017), the function of citations in the text (e.g., Ding et al., 2014; Garfield, 1965; Lyu et al., 2021; Petrić, 2007), and the tone or sentiment of citations (e.g., Aljuaid et al., 2021; Athar & Teufel, 2012; Vyas et al., 2020; Yousif et al., 2019). It has also led to studies proposing more fine-grained approaches to measure the impact of papers and studies focusing on the identification of essential references and ignoring non-essential or perfunctory citations (e.g., Moravcsik & Murugesan, 1975) that do not directly contribute to the substantive argument or analysis of the paper in which they are included. For example, Zhao and Strotmann (2020) introduced a location-filtered citation counting method with the aim of making essential citations more noticeable. Pak et al. (2020) proposed a fractional counting method based on in-text citations and argued that not all references that are mentioned only once are perfunctory. Lyu et al. (2021) argued that the value of a citation (e.g., importance and impact) should be dependent on the motivation of the citation. Aljohani et al. (2021) used a machine learning approach to detect important citations and Aljuaid et al. (2021) tried to improve the detection of important citations using sentiment analysis.

## 2.2 From static to dynamic perspectives in citation analysis

Citing behavior is the subfield of citation analysis that studies the patterns and motivations behind the citation practices of researchers in scholarly papers. It seeks to understand the reasons and rationales authors have for citing specific works in their own papers. Two main theories of citing behavior have been developed in early studies, both of them situated within broader social theories of science. One is often denoted as the normative theory of citing behavior and the other as the social constructivist view of citing behavior (e.g., Baldi, 1998; Cozzens, 1989; Bornmann & Daniel, 2008). The normative theory of citing behavior, introduced by Merton (1973), conceptualizes citations as a symbolic acknowledgment of intellectual debt. This perspective argues that researchers adhere to a universal set of norms when selecting and citing references, prioritizing the intrinsic merit and relevance of cited works over personal biases. According to this view, citations serve as objective indicators of a work's value within the scientific community, reflecting its genuine contribution to knowledge. In contrast, the social constructivist perspective challenges the idea that citing practice is governed solely by internal norms. Advocates such as Gilbert (1977), Latour and Woolgar, 1979, and Knorr-Cetina (1981) argue that scientific knowledge is shaped by social dynamics and external influences. From this viewpoint, citations are not just acknowledgments but also rhetorical tools that authors use to strengthen their arguments and persuade readers. Social constructivists highlight how factors like an author's social position, institutional affiliations, or strategic goals can influence citation choices, suggesting that these decisions are as much about negotiation and persuasion as they are about intellectual merit. The two theories of citing behavior provide distinct but complementary insights into why researchers cite particular works, and they align well with findings of other scholars. For instance, Lyu et al. (2021) analyzed thirty-eight studies and identified two main categories of reasons behind citing behavior, summarized as scientific motivations and tactical motivations.

The temporal dimension is crucial when studying and understanding citing behavior. Researchers have increasingly recognized that time significantly influences the evolution of citing practices and the perceived value of scholarly works. For instance, Boyack et al. (2018) showed that both the average number of references and the average number of in-text citations have increased over time. These evolving citing practices, in combination with the exponential growth of the scholarly literature, impact the perceived value and influence of individual citations (e.g., Lin et al., 2023; Petersen et al., 2019). The function of a paper within the literature, as well as the motivations for citing a particular work, can also shift over time. New findings or insights might emerge, leading to a re-evaluation and recontextualization of older research. Early investigations into this phenomenon can be traced back to Burton and Kebler's (1960) application of the concept of "half-life" from physics to characterize the obsolescence function or impact decay of scholarly papers. Typically, as papers age, their annual citation rate tends to decline. This decline reflects a trend of diminishing engagement with older papers and a preference for more recent works (e.g., Aversa, 1985; Glänzel & Schoepflin, 1995; Wang et al., 2013). However, there are exceptions to this general trend. For instance, Gou et al. (2022) found that some papers might experience a "literature revival", characterized by a resurgence in citation activity after a period of obsolescence.

## 2.3 The changing role of highly cited papers

HCPs, that keep attracting citations over time, have also been an interesting source of study for researchers. While HCPs are often viewed as influential works, it is crucial to recognize that citation count alone does not fully capture their impact or value within the scholarly literature. Citations can stem from a variety of motivations, ranging from acknowledging foundational contributions to simply mentioning a concept in passing. Bonzi (1982) highlights this variability by noting that some HCPs serve as cornerstones for further research, while others receive only cursory mentions.

Understanding the true impact of HCPs requires a deeper look into full text of the citing papers. Lu et al. (2017) observed that citations to Hirsch's (2015) HCP on the h-index paper shifted from being predominantly located in the introduction, results, and discussion sections in the early years to being more concentrated in the introduction and methodology sections later on. This shift suggests a change in how the paper is being used and the evolving role of the h-index from a novel concept to an accepted and applied metric. Similarly, Otto et al. (2019) found that, in general, the likelihood of HCPs being cited in the methodology section increases with their age.

It is important to recognize that HCPs do not always receive in-depth engagement from citing authors. Small (1978) introduced the concept of "concept symbols" to describe how the function of HCPs and the engagement with these papers can evolve over time. Initially, these papers may be cited for their specific content and contributions. However, as they accumulate citations, they can become symbolic representations of concepts, methodologies, or research areas, cited without substantial engagement with their original content. Small (1978) found evidence for this phenomenon in chemistry, noting that many HCPs had "uniform or standardized usage and meaning". This implies that over time, citations to these papers may become less about their specific content and claims and more about their symbolic representation of a broader idea.

The shift from substantial to superficial engagement aligns with the notion of perfunctory citations (e.g., Moravcsik & Murugesan, 1975), as previously discussed. Perfunctory citations are often found in introductory sections and serve to acknowledge a paper's existence or provide general background information without critically evaluating its content or considering more recent advancements. Voos & Dagaev (1976) and Huang et al. (2021) confirmed this tendency, noting that many HCPs acquire a significant portion of their citations from perfunctory citations in mainly introductory sections. Thelwall (2019b) also found that perfunctory citations are most likely to occur in the introduction section of scholarly works, but also demonstrated that many HCPs primarily attract citations from the methodology section, consistent with findings by Otto et al. (2019).

In conclusion, examining where citations appear within a scholarly paper, how frequently they occur, and the surrounding text and citations offers critical insights into the evolving role and impact of cited works on other papers.

# 3. Data and methods

To analyze the changing role of cited papers over time, we focused on HCPs published sufficiently long ago. This approach ensured the availability of a substantial number of citations for the same papers and allowed us to systematically analyze temporal differences.  For this purpose, we used the Web of Science (WoS) database, a widely used bibliographic databases covering literature from a broad range of disciplines (e.g., Birkle et al., 2020; Pranckutė, 2021). Specifically, we used the in-house version of the WoS database available at the Centre for Science and Technology

Studies (CWTS) at Leiden University. This version of the WoS database includes the Science Citation Index Expanded, the Social Sciences Citation Index, and the Arts & Humanities Citation Index. We identified HCPs in WoS by searching for all papers that were (1) cited at least 1000 times, (2) published in 2000, (3) written in English, and (4) classified as articles. In this way, we retrieved 883 HCPs. In total, these HCPs were cited by 1,054,664 unique papers indexed in WoS and published in the time period 2000-2016.

To study in greater detail how the HCPs were cited, we collected the full text of the citing papers. We sourced full-text data from the Elsevier ScienceDirect corpus that was also used in previous studies (Boyack et al., 2018; Lamers et al., 2021) and that is hosted at CWTS. This corpus contains the full texts of nearly five million English-language research articles, short communications, and review articles published in approximately 3,000 Elsevier journals between 1998 and 2016. We successfully retrieved the full text of 220,335 citing papers, accounting for 20.9% of the total number of citing papers that can be found in WoS for the time period 2000-2016. Figure 1 shows that the number of citing papers is increasing every year. However, the percentage of citing papers for which we were able to obtain the full text from the Elsevier corpus remains consistently around 20% across all years. This consistently confirms that we were able to collect a representative sample of citing papers with a full text for each of the citing years.

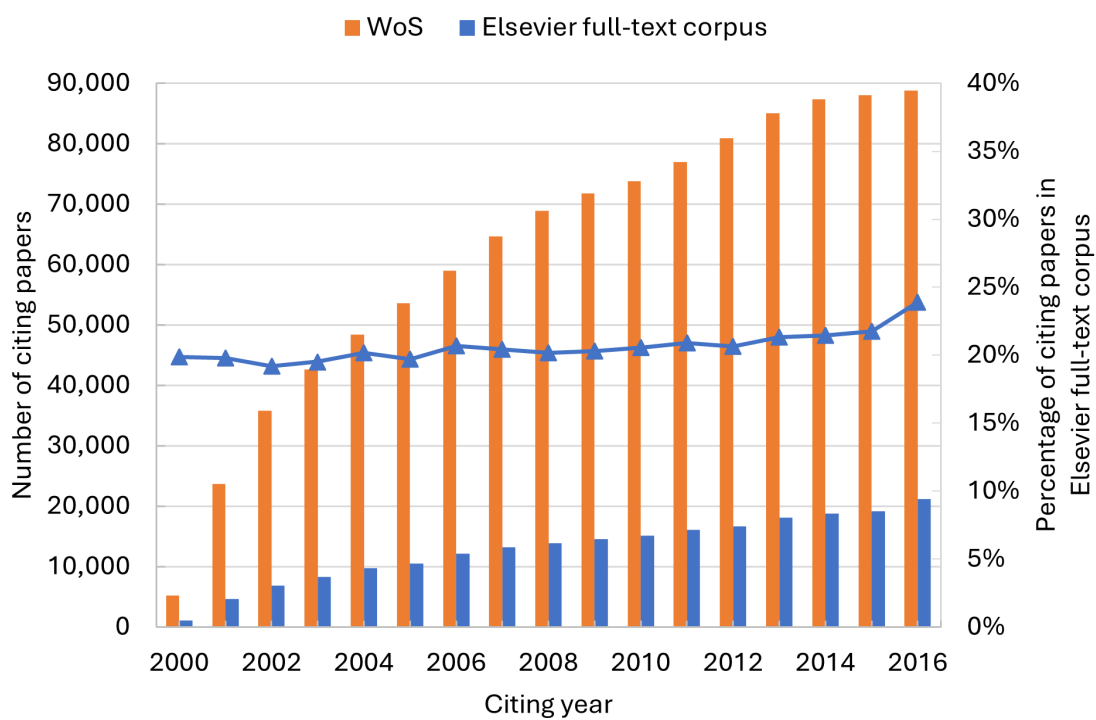

Figure 1. Number of citing papers and percentage of citing papers with a full text in different citing years.

The full text of the 220,335 papers citing our HCPs was parsed to identify references, citation sentences, in-text citations, and reference mentions. The obtained full-text and citation information of the 220,335 citing papers is summarized in the Table 1. Since HCPs can be cited together in the same citing paper, we identified a total of 251,645 references pointing to HCPs within the 220,335 citing papers. Additionally, because HCPs can be cited in the same in-text citation, the number of in-text citations is slightly lower than the number of reference mentions.

Table 1. The basic full-text and citation information of the citing papers.

|  | Referring to all cited works | Referring to the HCPs only |
|---|---|---|
| References | 12,259,405 | 251,645 |
| Reference mentions | 18,006,353 | 360,419 |
| In-text citations | 11,742,513 | 360,366 |
| Citation sentences | 9,638,642 | 357,099 |

After parsing the full text of the citing papers and identifying the references, citation sentences, in-text citations, and reference mentions in those papers, we determined various aspects of the citations to the HCPs, including citation location, frequency, sentiment, and the relatedness between the citing paper and the cited HCP. We also analyzed how these aspects change over time. The results of this analysis are presented in the next section.

# 4. Results

## 4.1 Citation location over time

We first analyzed the location of citations to the HCPs within the full text of the citing papers. Text progression was used to represent the citation location (e.g., Boyack et al., 2018). Text progression is expressed on a scale from 0 to 1 and is identified at the character level. Specifically, if an HCP is cited at the $i$th character in the full text of the citing paper and this paper has $n$ characters in total then the citation location is calculated as $i / n$. The average citation location per citing year was calculated using the arithmetic mean. Additionally, we divided the citation location based on the text progression into three equal parts: begin part (0%-33%), middle part (33%-66%), and end part (66%-100%).

Figure 2 shows that the average in-text location of citations to HCPs decreases over the citing years. This means that, on average, HCPs are cited earlier in the text as the difference in publication year between the citing papers and the HCPs increases. Figure 3 shows that while there is little change in the percentage of HCP in-text citations located in the middle part over the citing years (it fluctuates around 25%), the percentage located in the begin part is increasing and that in the end part is decreasing. This suggests that citations to HCPs in the first citing years are about equally common in the opening sections (36%), such as introduction or background, and the final sections (37%), such as discussion or conclusion. However, in later citing years this changes significantly, with more citations appearing in the opening sections (56%) and fewer in the final sections (19%).

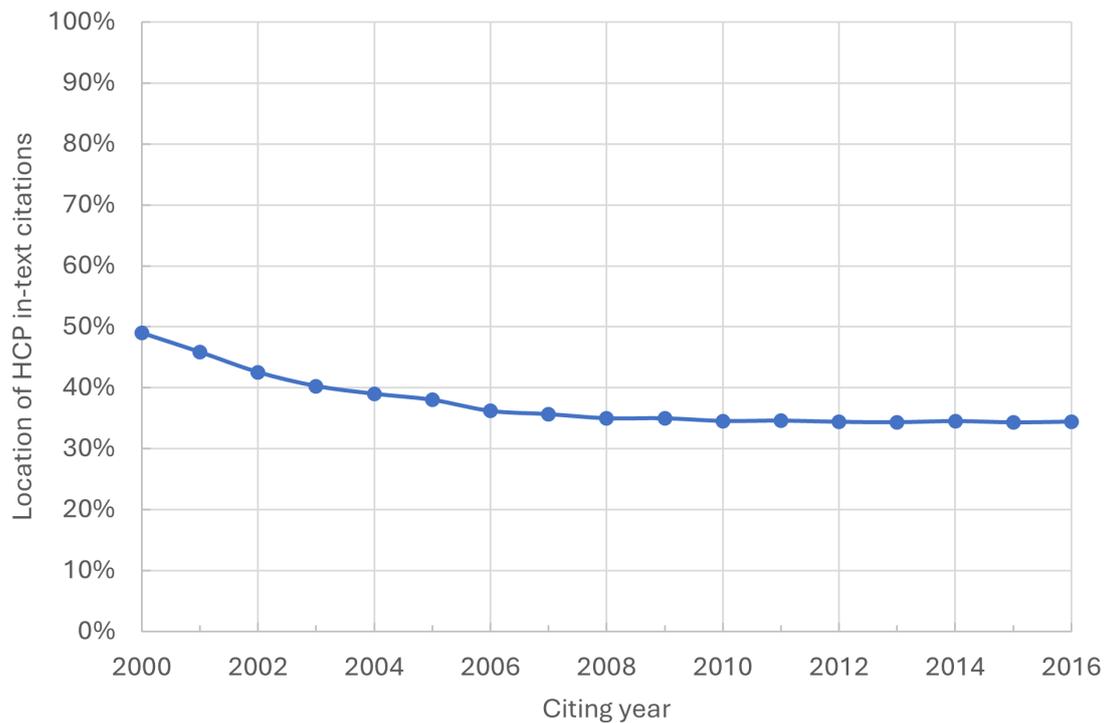

Figure 2. Average location of citations to HCPs within the full text of citing papers across different citing years.

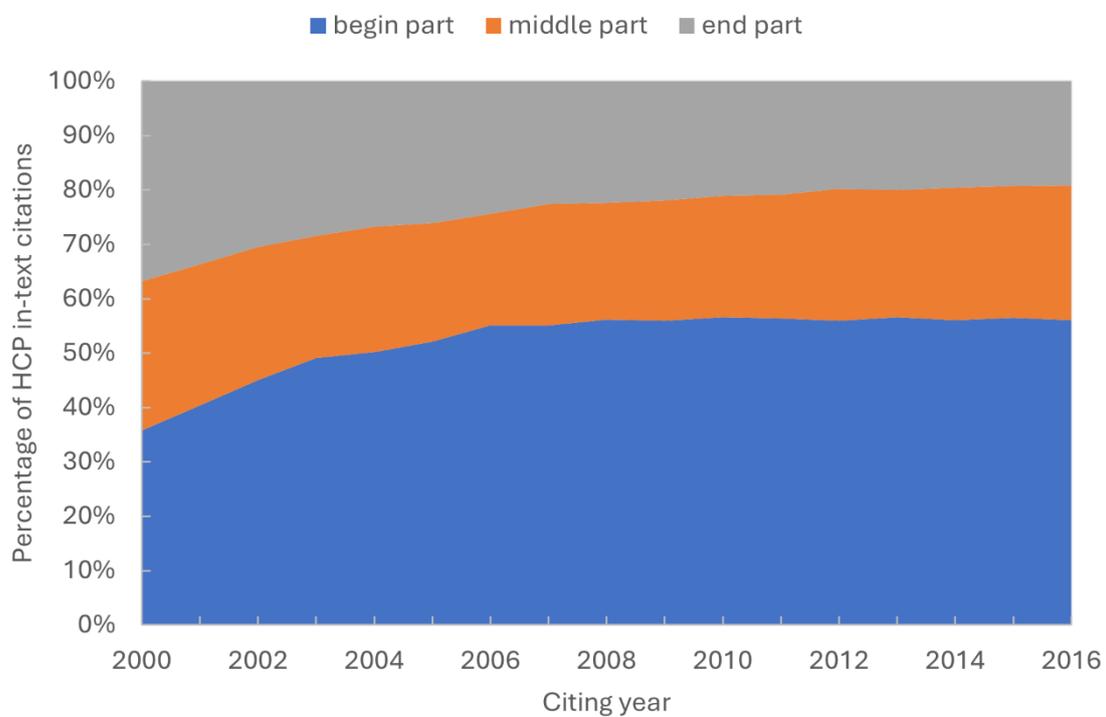

Figure 3. Percentage of citations to HCPs located in the begin part, middle part, and end part of the full text of citing papers across different citing years.

## 4.2 Cited reference type over time

References can be mentioned more than once in the full text. We distinguish between two types of references:

- Multiple Mentioned Reference (MMR): an MMR is a reference that is mentioned more than once in the body of the full text of the citing paper (Hu et al., 2017). It is also referred to as "CountX mentions" (Ding et al., 2013) or "Multi-citations" (Zhao et al., 2017).
- Single Mentioned Reference (SMR): an SMR is a reference that is mentioned only once in the body of the citing paper. It is also referred to as "CountOne mentions" (Ding et al., 2013) or "Uni-citations" (Zhao et al., 2017).

An increasing number of scholars argue that MMRs are more essential to the citing work than SMRs (Ding et al., 2013; Hu et al., 2017; Zhao & Strotmann, 2020). In this study, we analyzed the number of times references to HCPs are mentioned within the full text and examined how the percentages of SMRs and MMRs change over the citing years.

Figure 4 shows that, in the initial citing years, many HCPs are cited multiple times within the citing papers. However, as the citation interval between the citing papers and the HCPs increases, the percentage of MMRs decreases from 37% to 25% and the percentage of SMRs increases from 63% to 75%. It is also useful to know the average number of times that each reference is mentioned. Figure 5 provides insight into the average number of times each reference is mentioned. It shows that the average number of times an HCP is mentioned in the full text of the citing papers has decreasing from 1.8 in 2000 to 1.4 in 2016. This result is in line with previous findings suggesting that references mentioned only once tend to be more highly cited than those that are mentioned multiple times (Boyack et al., 2018).

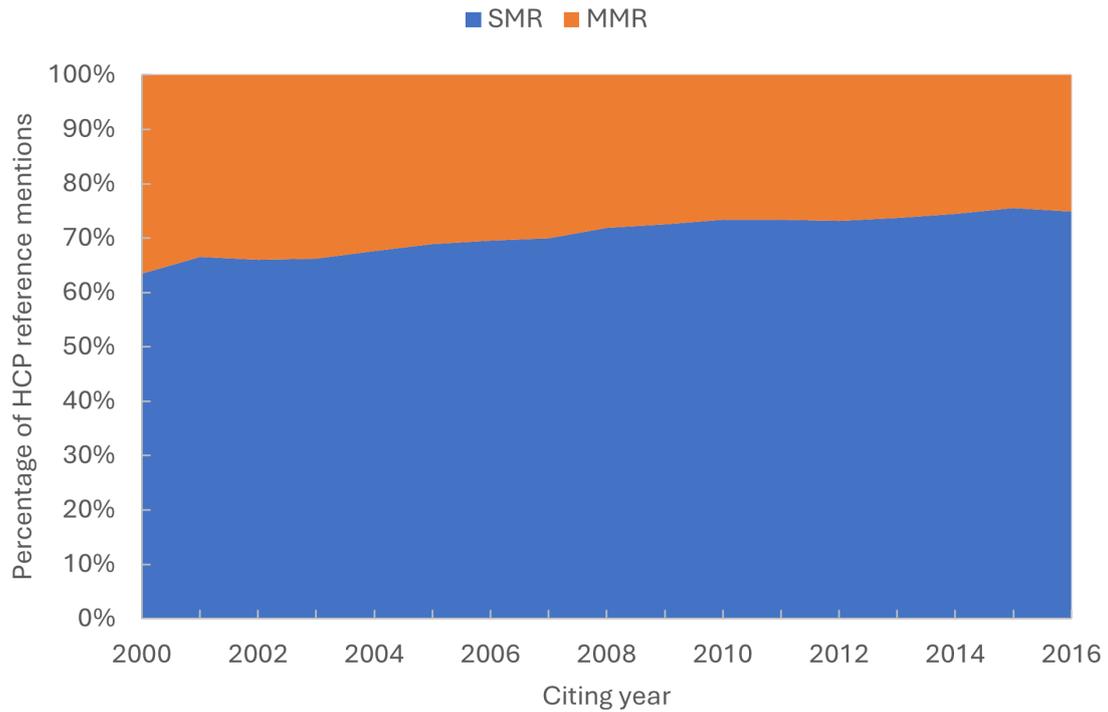

Figure 4. Percentage of references to HCPs that are mentioned only once (SMR) or multiple times (MMR) within the full text of citing papers across different citing years.

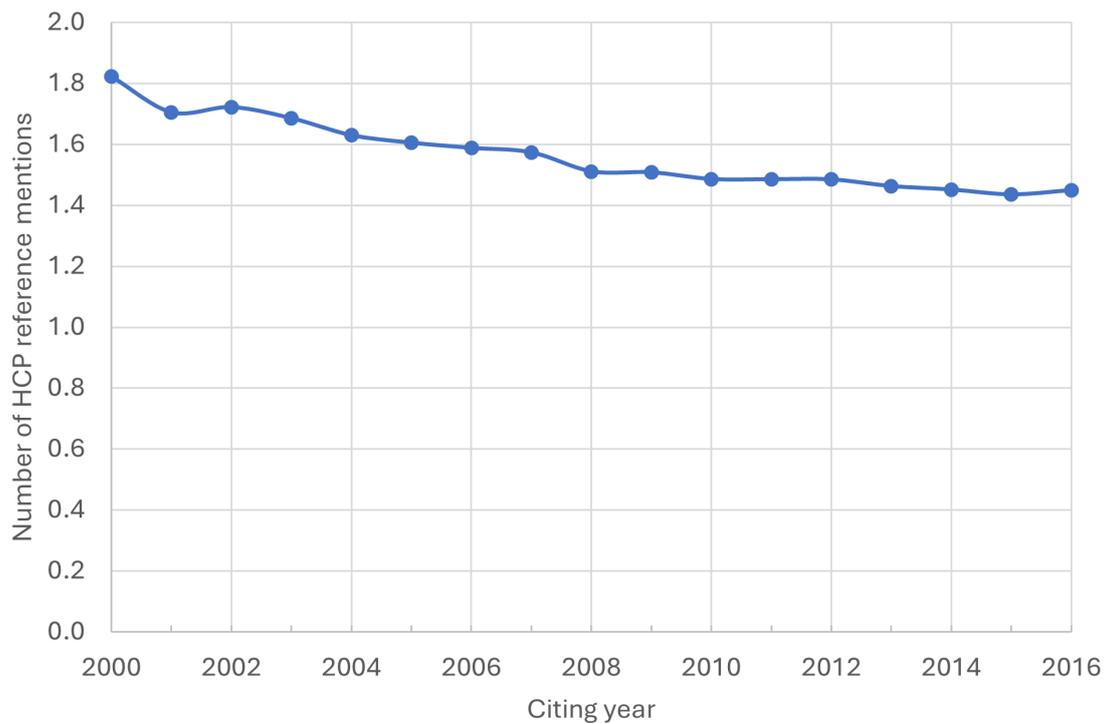

Figure 5. Average number of mentions of references to HCPs within the full text of citing papers across different citing years.

## 4.3 In-text citation type over time

In this study, we also analyzed the in-text citations in which the HCPs are referenced. In-text citations can contain one or more references. We distinguish between two types of in-text citations:

- Multi-Reference Citation (MRC): an MRC is an in-text citation that includes more than one reference (American Psychological Association, 2019; Lin et al., 2019). It is also referred to as "citation string" (Lyu et al., 2021), "citation co-mention" (Chao Lu et al., 2017), or "non-independent mention" (Huang et al., 2022; Pak et al., 2020).
- Single-Reference Citation (SRC): an SRC is an in-text citation that includes only one reference. It is also referred to as "independent mention" (Huang et al., 2022; Pak et al., 2020).

We analyzed the number of references that these in-text citations contain and how the percentage of SRCs and MRCs changes over the citing years. Figure 6 shows that the percentage of MRCs increases over the citing years, from 32.7% in 2000 to 45.7% in 2016. This suggests that, as HCPs age, they are more likely to be cited along with other references in the same in-text citation. This shift may indicate that as HCPs get older, they tend to serve more and more as general references and become less essential to the papers in which they are cited. Previous research has therefore warned against citation impact formulas that overvalue HCPs by treating their citations equally to those of less-cited papers (Thelwall, 2019b).

Figure 7 shows the average number of references in in-text citations that include a reference to an HCP. The number increases rapidly in the first few citing years, peaks in 2007, and then slightly declines through 2016. In other words, as HCPs age, they tend to be cited alongside more references, but this increase levels off after a few years.

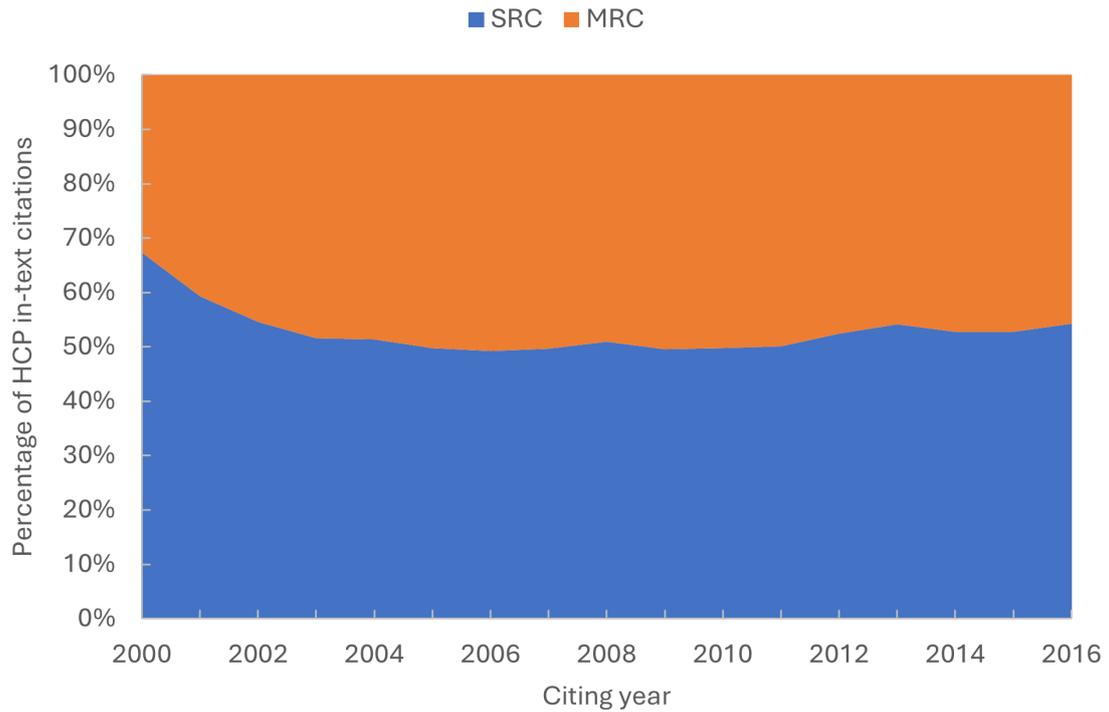

Figure 6. Percentage of in-text citations to HCPs that include only one reference (SRC) or multiple references (MRC) across different citing years.

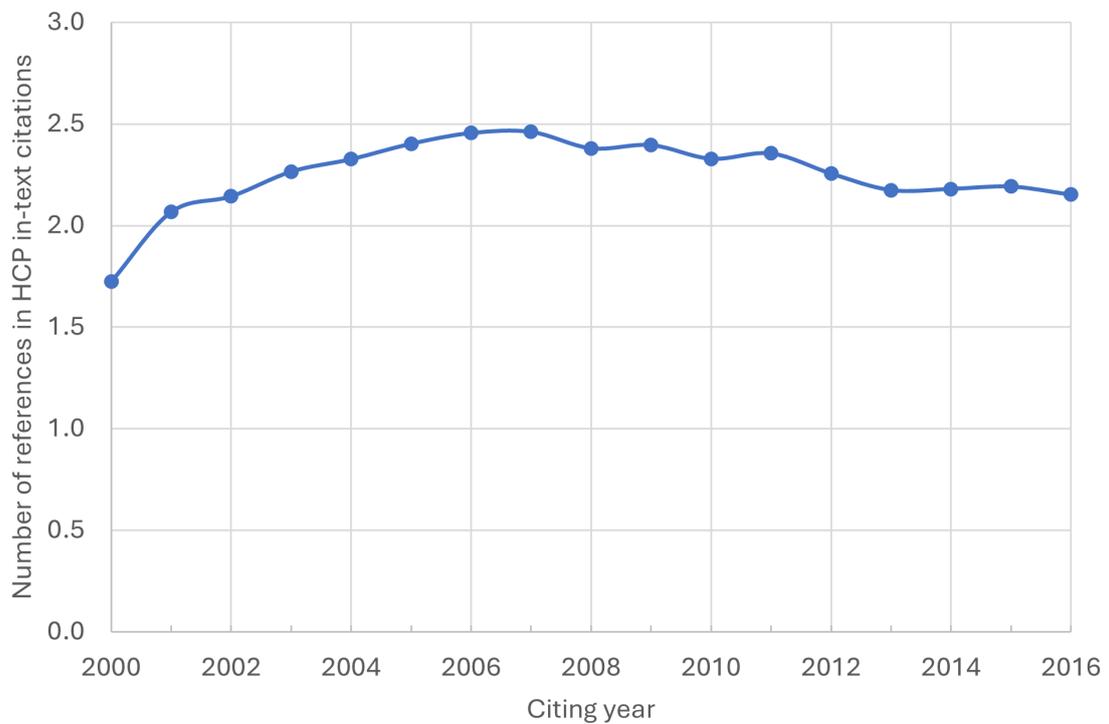

Figure 7. Average number of references in in-text citations to HCPs across different citing years.

## 4.4 Citation sentiment over time

Citation sentiment analysis, an important component of citation context analysis, is used to detect the emotional tone or attitude expressed by authors towards the works they cite (e.g., Vyas et al., 2020). Unlike traditional citation counts, this approach examines the context surrounding each in-text citation to determine whether the citing author expresses agreement, disagreement, neutrality, or other sentiments toward the cited work. Citation sentiment analysis offers valuable insights into the influence and contribution of papers, helps identify influential papers, and highlights areas of consensus or debate within a field.

VADER (Valence Aware Dictionary and sEntiment Reasoner) is a lexicon and rule-based sentiment analysis tool that determines, for each given text, the proportion of the text that is associated with a positive, negative, or neutral sentiment (Hutto & Gilbert, 2014). VADER is available in the NLTK package for Python and we used it to analyze how the sentiment of sentences containing citations to our HCPs changes over time.

Figure 8 shows that the largest proportion of the text of HCP citation sentences is associated with a neutral sentiment, consistently exceeding 90% across all citing years. The proportion associated with a positive sentiment has increased slightly over time, ranging from 4.7% to 6.1%. Meanwhile, the proportion associated with a negative sentiment has shown a marginal decline, decreasing from 3.8% to 3.5%.

VADER also provides a compound score, a 'normalized, weighted composite score', that offers a single, unidimensional measure of sentiment for a given text. The compound score ranges from -1 (most extreme negative) to +1 (most extreme positive). Figure 9 shows the average sentiment compound score of sentences containing citations to our HCPs. The results indicate a slight rise in sentiment as HCPs age, increasing from 0.04 to 0.11.

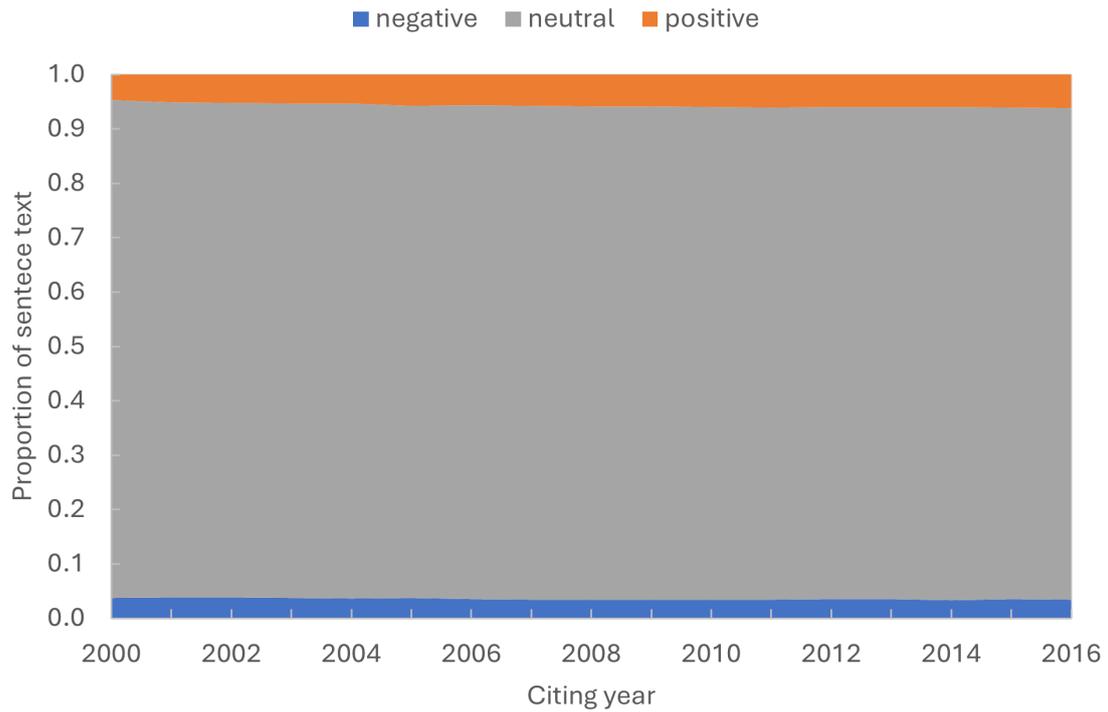

Figure 8. Proportion of the text of sentences containing citations to HCPs associated with a positive, negative, or neutral sentiment over the citing years.

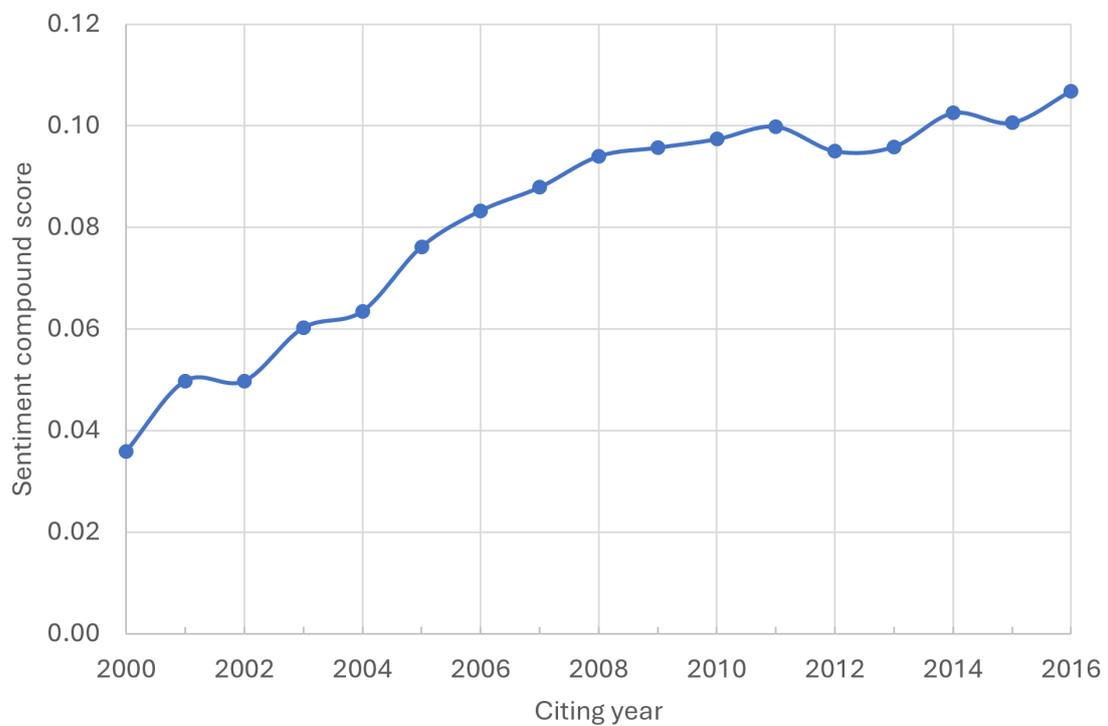

Figure 9. Average sentiment compound score of the text of sentences containing citations to HCPs over the citing years.

## 4.5 Relatedness over time

Given that the HCPs examined in our analysis are cited in the citing papers, it is reasonable to expect some degree of relatedness between the HCPs and their citing papers. However, papers can be cited for different reasons and may serve various functions in a citing paper (e.g., Lyu et al., 2021). For instance, an HCP might be cited because it addresses the same topic, uses a similar methodology, or serves as background information. An interesting question is whether the degree of relatedness changes as HCPs age and accumulate more citations. We therefore finally analyzed the relatedness between HCPs and their citing papers and how this relatedness evolves over time.

We used two approaches to measure relatedness between the HCPs and their citing papers. The first approach focused on textual relatedness, calculated using the titles and abstracts of the HCPs and citing papers (e.g., Colavizza et al., 2018). The underlying idea of this approach is that a higher similarity in the titles and abstracts of an HCP and its citing paper suggests a stronger topical relationship. To compute this, we used SciBERT (Beltagy et al., 2019), a BERT-based language model trained on scientific text (Devlin et al., 2019), to transform the titles and abstracts of both the HCPs and citing papers into embedding vectors. We then calculated the cosine similarity between the embeddings of the HCPs and their citing papers.

Figure 10 shows that the average textual relatedness between HCPs and their citing papers first increases slightly and then gradually decreases over the citing years. This indicates that the greater the difference in publication year between the HCP and its citing papers, the less closely related they are in terms of the language they use in their title and abstract.

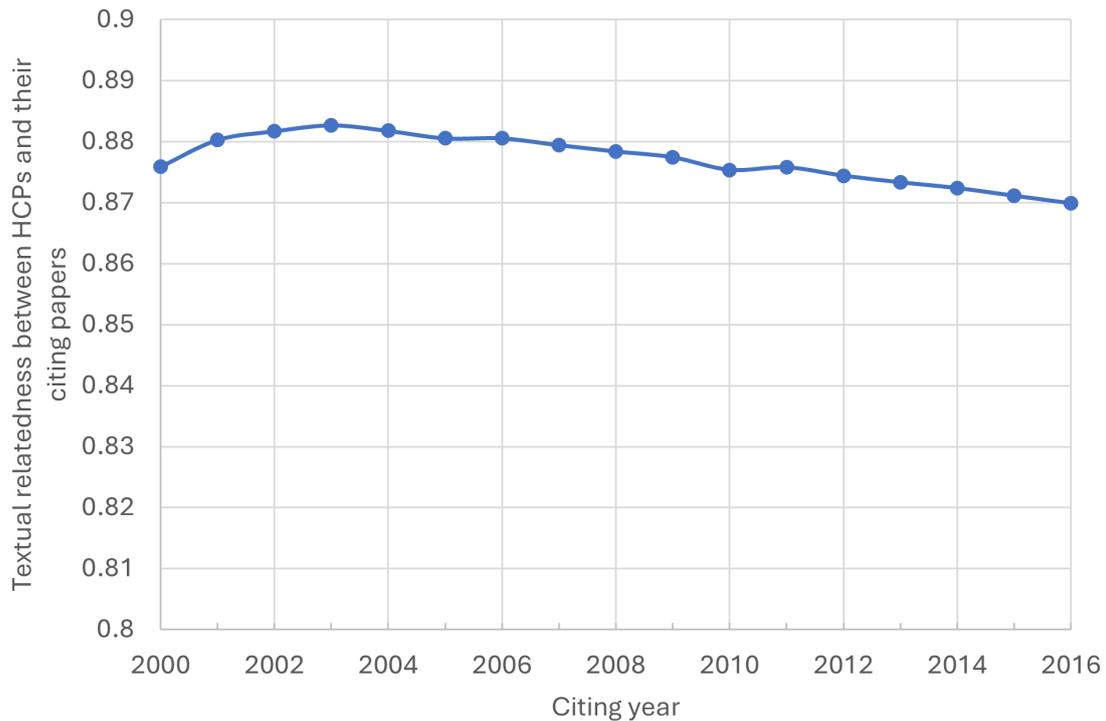

Figure 10. Average textual relatedness between HCPs and their citing papers.

In addition to textual relatedness, we also used another approach to calculate the relatedness between the HCPs and their citing papers, focusing on the references that they share. This approach, known as bibliographic coupling (Kessler, 1963), is based on the idea that the more references an HCP and a citing paper have in common, the more likely the two papers are related to each other. To account for differences in the length of reference lists, we used the Ochiai coefficient (i.e., cosine similarity applied to binary data, e.g., Van Eck & Waltman, 2009) to calculate the reference relatedness between HCPs and their citing papers.

Figure 11 shows that the average reference relatedness between HCPs and their citing papers decreases steadily over the citing years, from 0.09 in 2000 to 0.02 in 2016. The decrease is strongest in the first five years and gradually weakens thereafter. This indicates that the greater the age difference between the HCP and its citing paper, the less closely related they are in terms of the literature they cite.

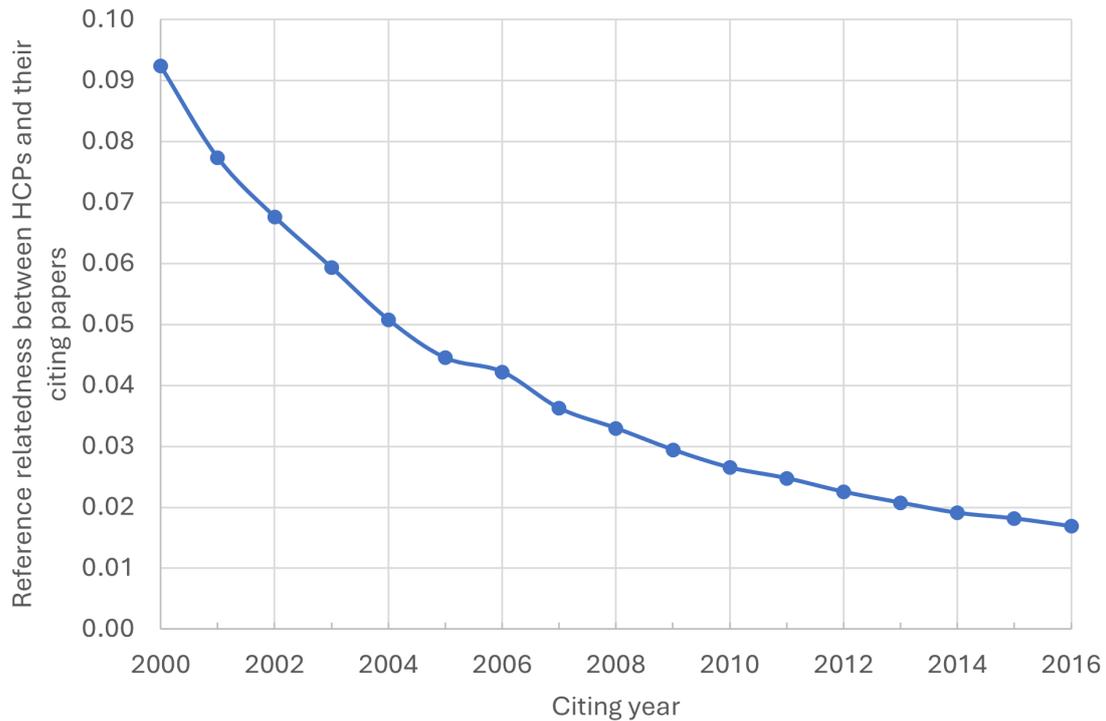

Figure 11. Average reference relatedness between HCPs and their citing papers.

The textual relatedness results (Figure 10), together with the reference relatedness results (Figure 11), suggest that on average HCPs become less closely related to the papers in which they are cited as they get older. This may indicate a shift in the function or role of HCPs and the reason why they are cited: as HCPs become more highly cited, they may increasingly serve as general background or symbolic references rather than as sources of direct topical or methodological relevance.

# 5. Discussion and conclusions

In this paper, we have analyzed how the role of cited papers changes over time. We have done this by focusing on a set of nearly 900 HCPs published between 2000 and 2016 and by examining the full text of more than 220 thousand papers citing these HCPs. Using this large full-text dataset, we have investigated how various aspects of citations to the HCPs change over time. The aspects included in our analysis are the location of the citation in the full text, cited reference type, in-text citation type, citation sentiment, and textual and bibliographic relatedness between the citing paper and the cited HCP.

The findings of our analysis can be summarized as follows. First, as HCPs age, they are on average cited earlier in the full text of their citing papers. The proportion of citations appearing in the middle sections remains relatively stable, while citations in the opening sections increase and those in the concluding sections decrease. Second, HCPs initially tend to be mentioned multiple times in the full text of their citing papers, but the proportion of citing papers for which this is the case declines as the HCPs age. The average number of times HCPs are mentioned in the full text of their citing papers also decreases over time. Third, older HCPs are more likely to be cited alongside other references within the same in-text citation, indicating they increasingly serve as general references rather than essential sources to the papers in which they are cited. Fourth, citation sentiment shows only a very weak increase in positivity over the citing years. By far, most of the text in sentences citing HCPs is associated with neutral sentiment, followed by positive and then negative sentiment. Finally, relatedness between HCPs and their citing papers decreases as HCPs age, both in textual similarity based on titles and abstracts, and in shared references.

Taken together, these patterns indicate a gradual shift in the function of HCPs as they become older and more highly cited. Initially, they tend to be engaged with more directly, discussed in greater detail, and more closely connected to the citing paper. Over time, however, they are more likely to serve as general, background, or symbolic references, signaling recognition and authority rather than direct topical or methodological influence. These results reflect average trends and do not necessarily capture the trajectory of any individual paper. Moreover, an HCP that is, on average, cited for symbolic reasons may still play a crucial role in the arguments, methods, or results of certain citing papers.

Our findings have implications for bibliometric research and research evaluation practices. Citation counts alone can obscure shifts in the role and influence of a work over time, as older HCPs may receive citations largely for symbolic or background purposes. Understanding the contexts of citations and their evolving nature can help refine measures of scientific impact, guide the interpretation of citation-based indicators, and improve the mapping of science by revealing more nuanced connections between papers. Moreover, the observed decline in relatedness between HCPs and their citing papers over time indicates the need to distinguish between citations signaling direct intellectual debt and those serving primarily as markers of prestige or historical significance. Importantly, our analysis shows that the full text of citing papers contains valuable clues about these changes and can be systematically leveraged to identify shifts in the role of cited works over time. This highlights the importance of full-text–based approaches in advancing bibliometric research and developing more nuanced measures of scholarly influence.

There are several limitations to this study that should be noted. Although the number of full texts that we have collected is large, it still accounts for only a relatively modest share of the total number of citing papers (14.3%). A more comprehensive picture could be obtained by incorporating additional full-text data from other sources. Furthermore, we have not analyzed disciplinary or field-specific patterns, which may have concealed evolutionary features present at more granular levels. We look forward to future studies that examine the evolutionary characteristics of citations, at more granular levels, using full-text data from multiple sources, considering different research areas, and using advanced semantic approaches. Such studies can deepen our understanding of citation theory and behavior, and inform practical applications in information retrieval, scholarly search, and the modeling of the structure and dynamics of science.

## Data availability

The Elsevier full-text data and the Web of Science bibliographic data used in this study were obtained from proprietary data sources. We are not allowed to share the raw data on which our study is based. The data underlying the statistics reported in the table and figures of this paper is openly available in Zenodo (Lin & Van Eck, 2025).

## Code availability

The source code used in the analysis of this paper is available in the following GitHub repository: https://github.com/neesjanvaneck/changing-role-cited-papers.

## Author contributions

Gege Lin: Conceptualization, Methodology, Formal analysis, Writing— original draft. Nees Jan van Eck: Conceptualization, Methodology, Formal analysis, Writing— review & editing. Haiyan Hou: Conceptualization. Zhigang Hu: Conceptualization.

## Competing interests

The authors have no competing interests.

# Funding information

This work was supported by the National Natural Science Foundation of China (grant no. 71974030 and 72504017) and the 2022 Economic and Social Development Research Project of Liaoning Province, Liaoning Provincial Federation Social Science Circles (grant no. 2022lslybkt-034).